\documentclass[12pt,a4paper]{iopart}
\usepackage{amssymb,graphicx}
\newcommand{\avg}[1]{\langle #1\rangle}
\newcommand{\var}{\mathrm{var}}
\newcommand{\N}{\mathcal{N}}
\newcommand{\dd}{\mathrm{d}}
\begin{document}
\title{Breakdown of the mean-field approximation in a~wealth
distribution model}
\author{M Medo$^{1,2}$}
\address{$^1$ Physics Department, University of Fribourg,
P\'erolles, 1700~Fribourg, Switzerland}
\address{$^2$ Department of Mathematics,
Physics and Informatics, Mlynsk\'a dolina,
842~48~Bratislava, Slovak republic}
\ead{matus.medo@unifr.ch}
\begin{abstract}
One of the key socioeconomic phenomena to explain is the
distribution of wealth. Bouchaud and M\'ezard have proposed an
interesting model of economy [Bouchaud and M\'ezard (2000)]
based on trade and investments of agents. In the mean-field
approximation, the model produces a stationary wealth
distribution with a power-law tail. In this paper we examine
characteristic time scales of the model and show that for any
finite number of agents, the validity of the mean-field result
is time-limited and the model in fact has no stationary wealth
distribution. Further analysis suggests that for heterogeneous
agents, the limitations are even stronger. We conclude with
general implications of the presented results.
\end{abstract}
\pacs{05.40.-a, 89.65.-s, 89.75.-k
Submitted to: JSTAT}

\maketitle

\section{Introduction}
Many empirical studies report broad distributions of income and
wealth of individuals and these distributions are often claimed
to have power-law tails with exponents around two for most
countries \cite{Pareto,Pig84,Ao00,DrYak01,Sinha06}. The first
models attempting to explain the observed properties appeared
over fifty years ago \cite{Champ53,WW57,Stig69}. Much more
recently, physics-motivated kinetic models based on random
pairwise exchanges of wealth by agents have attracted
considerable interest \cite{Isp98,DrYak00,Sla04,Pat06,ChCh07}.
An alternative point of view is adopted in the wealth
redistribution model (WRM) where agents continuously exchange
wealth in the presence of noise \cite{BM00,Sorin01,MAH04}. There
are also several specific effects which can lead to broad wealth
distributions \cite{Sorn98,HuSo01,Reed01}. (For reviews of power
laws in wealth and income distributions
see~\cite{QRR97,DS00,Yak07}, while for general reviews of power
laws in science see \cite{New05,Farm06}.)

In this paper we analyze the WRM with two complementary goals in
mind. Firstly we investigate the simplest case when exchanges of
all agents are identical, focusing on the validity of the
mean-field approximation which is the standard tool to solve the
model and derive  the stationary wealth distribution. In
particular, we show that for any finite number of agents there
is no such stationary distribution (other finite-size effects
are discussed for a~similar model in~\cite{HuSo01}). Secondly we
investigate the model's behaviour when the network of agent
exchanges is
heterogeneous. Previous attempts to investigate the influence of
network topology on the model~\cite{BM00,Sou01,GaLo04,GaLo08}
were all based on the mean-field approximation. We show that
this is questionable because heterogeneity of the exchange
network strongly limits the validity of results obtained using
the mean-field approximation.

\section{Model and its mean field solution}
Adopting the notation used in~\cite{BM00}, we study a simple
model of an economy which is composed of $N$ agents with wealth
$v_i$ ($i=1,\dots,N$). The agents are allowed to mutually
exchange their wealth (representing trade) and they are also
subject to multiplicative noise (representing speculative
investments). The time evolution of agents' wealth is given by
the system of stochastic differential
equations (SDEs)
\begin{equation}
\label{dv-general}
\dd v_i(t)=\Big(\sum_{j\neq i}J_{ij}v_j(t)-
\sum_{j\neq i} J_{ji}v_i(t)\Big)\,\dd t+
\sqrt2\sigma v_i(t)\,\dd W_i(t),
\end{equation}
where $\sigma\geq0$ controls the noise strength. The coefficient
$J_{ij}$ quantifies the proportion of the current wealth
$v_j(t)$ that agent $j$ spends on the production of agent $i$
per unit time. We assume the It\^o convention for SDEs and
$\dd W_i(t)$ is standard white noise~\cite{Gard04,VanK07}. Hence,
denoting averages over realisations by $\avg{\cdot}$, we have
$\avg{\dd W_i(t)}=0$,
$\avg{\dd W_i(t)\,\dd W_j(t)}=\delta_{ij}\,\dd t$, and
$\avg{v_i(t)\,\dd W_i(t)}=0$. By summing $\dd v_i(t)$ over all
agents one can see that the average wealth
$v_A(t):=\case1N\sum_{i=1}^N v_i(t)$ is not influenced by wealth
exchanges and obeys the SDE
$\dd v_A(t)=\case{\sqrt2\sigma}N\sum_{i=1}^N v_i(t)\,\dd W_i(t)$.
Therefore $\avg{\dd v_A(t)}=0$ and $\avg{v_A(t)}$ is constant.
For simplicity we assume $v_i(0)=1$ ($i=1,\dots,N$) and thus
$\avg{v_i(t)}=1$ and $\avg{v_A(t)}=1$. (The influence of the
initial conditions is discussed in \Sref{sec:initial}.)

The system behaviour is strongly influenced by the exchange
coefficients $J_{ij}$. The simplest choice is $J_{ij}=J/(N-1)$
where all exchanges are equally intensive---we say that the
exchange network is homogeneous. By rescaling the time we can
set $J=1$ which means that during unit time agents exchange all
their wealth. Consequently, \eref{dv-general} simplifies to
\begin{equation}
\label{dv-homo}
\dd v_i(t)=\big(\tilde v_i(t)-v_i(t)\big)\,\dd t+
\sqrt2\sigma v_i(t)\,\dd W_i(t)
\end{equation}
where $\tilde v_i(t):=\case1{N-1}\sum_{j\neq i}v_j$ is the
average wealth of all agents but agent $i$. In the limit
$N\to\infty$, fluctuations of $\tilde v_i(t)$ are negligible and
one can replace $\tilde v_i(t)\to\avg{\tilde v_i(t)}=1$ as
in~\cite{BM00}. Agents then effectively interact only with the
``mean field'' and their wealth levels are independent. Using
the Fokker-Planck equation for the wealth distribution
$f(v_i,t)$, the stationary solution $f(v_i)$ can be found in the
form
\begin{equation}
\label{fv-BM}
f(v_i)=\frac{(\lambda-1)^{\lambda}}{\Gamma(\lambda)}\,
\exp\Big[-\frac{\lambda-1}{v_i}\Big]v_i^{-1-\lambda},\quad
\lambda:=1+1/\sigma^2.
\end{equation}
For $v_i\gg\lambda-1$, $f(v_i)$ decays approximately as
a~power-law with exponent $2+1/\sigma^2$, while the cumulative
distribution has exponent $1+1/\sigma^2$. When $v_i$ is well
described by \eref{fv-BM}, we say that the system is in the
\emph{power-law regime}.

The empirical studies mentioned above report power-law exponents
around $2$, indicating that in this model, $\sigma\simeq1$ is
needed to obtain realistic power-law behaviour of the wealth
distribution. In our analytical calculations we assume
$\sigma<1$; strong noise ($\sigma\geq1$) is discussed separately
at the end of the following section.

\section{Complete exchange network for a finite $N$}
To examine when the power-law regime is realised, we first
investigate the time needed to reach the mean-field solution
\eref{fv-BM}. Such relaxation times were studied very recently in
kinetic models of wealth distribution~\cite{Gup08}.

Given the homogeneous initial conditions $v_i(0)=1$
($i=1,\dots,N$), the exchange terms proportional to
$\tilde v_i-v_i$ are zero at $t=0$ and can be neglected for
small times. Hence when $t$ is small, each $v_i(t)$ evolves
independently due to multiplicative noise, $v_i(t)$ is
lognormally distributed, and its variance is
$\var[v_i](t)=\exp[2\sigma^2t]-1=2\sigma^2t+O(t^2)$: we say that
the system is in the \emph{free regime}. From the known variance
$\sigma^2/(1-\sigma^2)$ of the mean-field solution \eref{fv-BM},
we can estimate the transition time $t_1$ between the free
regime and the power-law regime as
\begin{equation}
\label{t1}
t_1=\frac1{2(1-\sigma^2)}.
\end{equation}
When $t\gg t_1$, the system has been given enough time to reach
the power-law regime.

We now recall the average wealth $v_A(t)$. While
$\avg{\dd v_A(t)}=0$, one can see that $\avg{\dd v_A^2(t)}$ is
always positive. Hence the variance $\var[v_A(t)]$ grows without
limit, in contradiction with the variance of \eref{fv-BM} which
is finite for $\sigma<1$. To resolve this disagreement we have
to accept that $f(v_i)$ as given by \eref{fv-BM} is not
a~stationary solution. But what comes after the power-law
regime? Since the Fokker-Planck equation for the joint
probability distribution $f(v_1,\dots,v_N)$ cannot be solved
analytically, we answer this question by investigating the
average quantities $\avg{v_i^2(t)}$ and $\avg{v_i(t)v_j(t)}$
($i\neq j$); now we are considering $\sigma<1$ and hence both
are well defined. Due to the assumed homogeneous network of
interactions and the chosen initial conditions, all averages
$\avg{v_i^2(t)}$ are identical and the same holds for the
cross-terms $\avg{v_i(t)v_j(t)}$; effectively we are left with
only two variables. From the It\^o lemma it follows that
$\dd(v_i^2)=(2v_i+\dd v_i)\,\dd v_i$ and
$\dd(v_iv_j)=v_i\,\dd v_j+v_j\,\dd v_i+\dd v_i\,\dd v_j$. After
substitution of \eref{dv-homo} and averaging over all possible
realisations, we obtain the exact set of equations
\begin{equation}
\label{evolution-CN}
\eqalign{
\frac{\dd\avg{v_i^2(t)}}{\dd t}&=2\,\Big[
\avg{v_i(t)v_j(t)}-(1-\sigma^2)\avg{v_i^2(t)}\Big],\\
\frac{\dd\avg{v_i(t)v_j(t)}}{\dd t}&=\frac2{N-1}\,\Big[
\avg{v_i^2(t)}-\avg{v_i(t)v_j(t)}\Big].}
\end{equation}
Since we set $v_i(0)=1$ ($i=1,\dots,N$), $\avg{v_i(t)}=1$ and
the initial conditions are $\avg{v_i^2(0)}=1$ and
$\avg{v_i(0)v_j(0)}=1$; for the general case see
\Sref{sec:initial}. Independently of the initial conditions, for
$\sigma>0$, \eref{evolution-CN} has only the trivial stationary
solution $\avg{v_i^2}=\avg{v_iv_j}=0$. This confirms that for
a~finite $N$, there is no stationary distribution $f(v_i)$.

By solving \eref{evolution-CN} one obtains the variance
$\var[v_i](t)=\avg{v_i^2(t)}-\avg{v_i(t)}^2$ as a function of
time and as a by-product also the correlation between agents
$i$ and $j$
\begin{equation}
\label{C}
C_{ij}(t):=\frac{\avg{v_i(t)v_j(t)}-\avg{v_i(t)}\avg{v_j(t)}}
{\sqrt{\var[v_i(t)]\var[v_j(t)]}}.
\end{equation}
Since the resulting expressions are rather complicated, here we
discuss only their limiting cases. Small time expansions can be
found in the form
\begin{eqnarray}
\label{var-smallt}
\var[v_i](t)&=&2\sigma^2\,t+O(t^2),\\
\label{C-smallt}
C_{ij}(t)&=&\case1{N-1}\,t+O(t^2).
\end{eqnarray}
As can be seen, \eref{var-smallt} agrees with our previous
reasoning about the log-normal nature of $f(v_i)$ in the free
regime, while \eref{C-smallt} confirms that in the limit
$N\to\infty$, wealth correlations vanish.

In the limit of large time we obtain
\begin{equation}
\label{C-larget}
\lim_{t\to\infty} C_{ij}(t)=
1-\sigma^2+\frac{\sigma^2}{1-\sigma^2}\,\frac1N+O(1/N^2).
\end{equation}
Thus, as $t$ increases, the system passes to the
\emph{synchronized regime} where the wealth of agents is
strongly correlated. One can estimate the transition time by
comparing the initial linear growth of $C_{ij}$ with its
stationary value, leading to
\begin{equation}
\label{t2}
t_2=(1-\sigma^2)\,N+O(1).
\end{equation}
An alternative estimate can be obtained from $\var[v_i](t)$.
Apart from a constant, it contains only terms proportional to
$\exp[\lambda_{1,2}t]$ where
\begin{equation*}
\lambda_{1,2}=\frac{-\sigma^2-N(1-\sigma^2)\pm
\sqrt{N^2(1-\sigma^2)^2+2N\sigma^2(3-\sigma^2)-
\sigma^2(4-\sigma^2)}}{2(N-1)}.
\end{equation*}
Since $\lambda_1<0$, $\lambda_2>0$, and for $\sigma\lesssim1$ is
$\vert\lambda_1\vert\gg\vert\lambda_2\vert$, the terms
proportional to $\exp[\lambda_1t]$ cause the initial saturation
of $\var[v_i](t)$ but the terms proportional to
$\exp[\lambda_2t]$ eventually take over and cause the divergence
of $\var[v_i](t)$. The corresponding transition time can be
roughly estimated by solving $\lambda_2t=1$, yielding
\begin{equation}
\label{t3}
t_3=\frac{1-\sigma^2}{2\sigma^2}\,N+O(1).
\end{equation}
Both $t_2$ and $t_3$ describe the transition between the
power-law and synchronized regimes: the former focuses on the
growth of correlations, the latter on the growth of variances.

\begin{figure}
\centering
\includegraphics[scale=0.28]{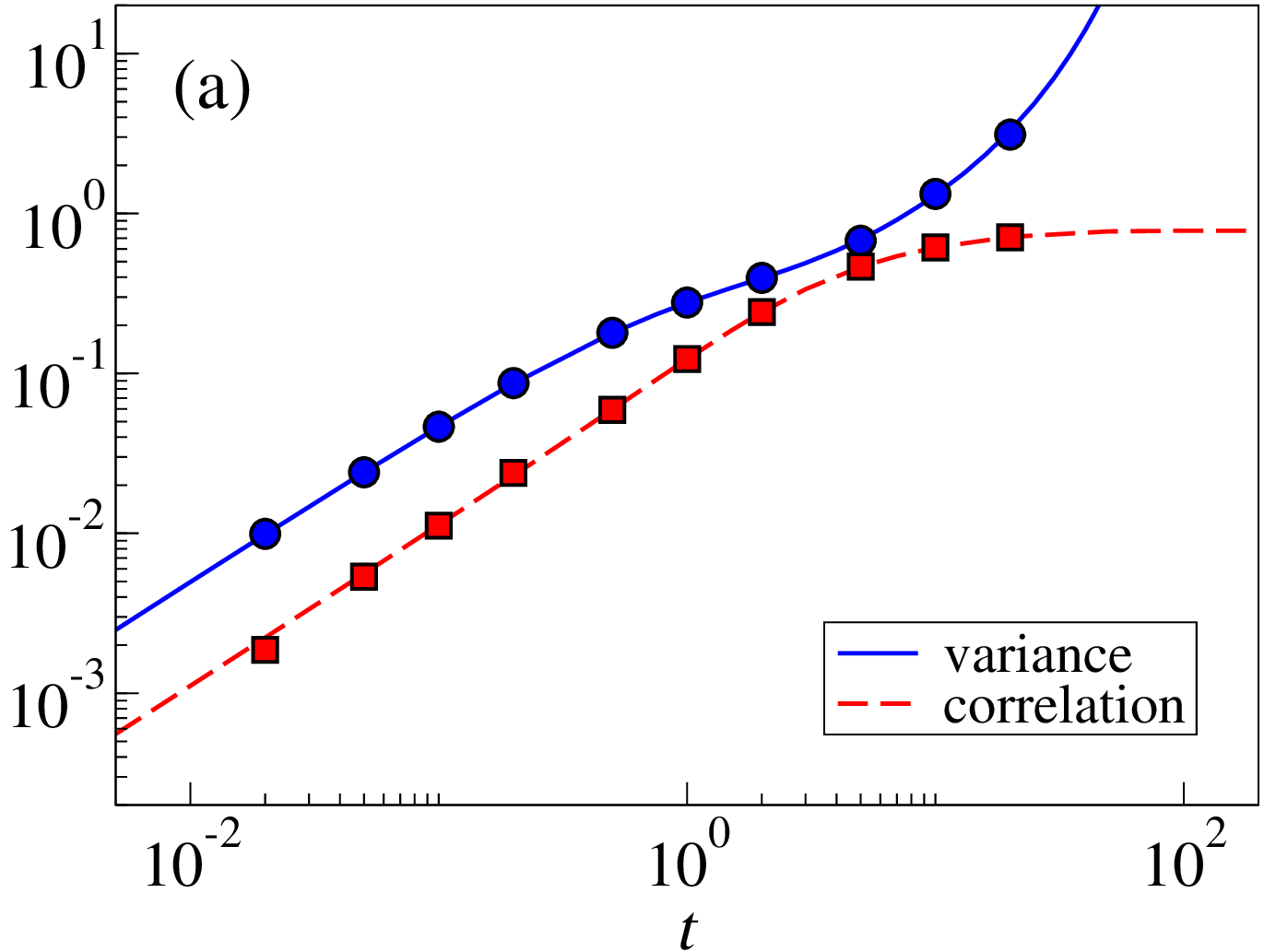}\qquad
\includegraphics[scale=0.28]{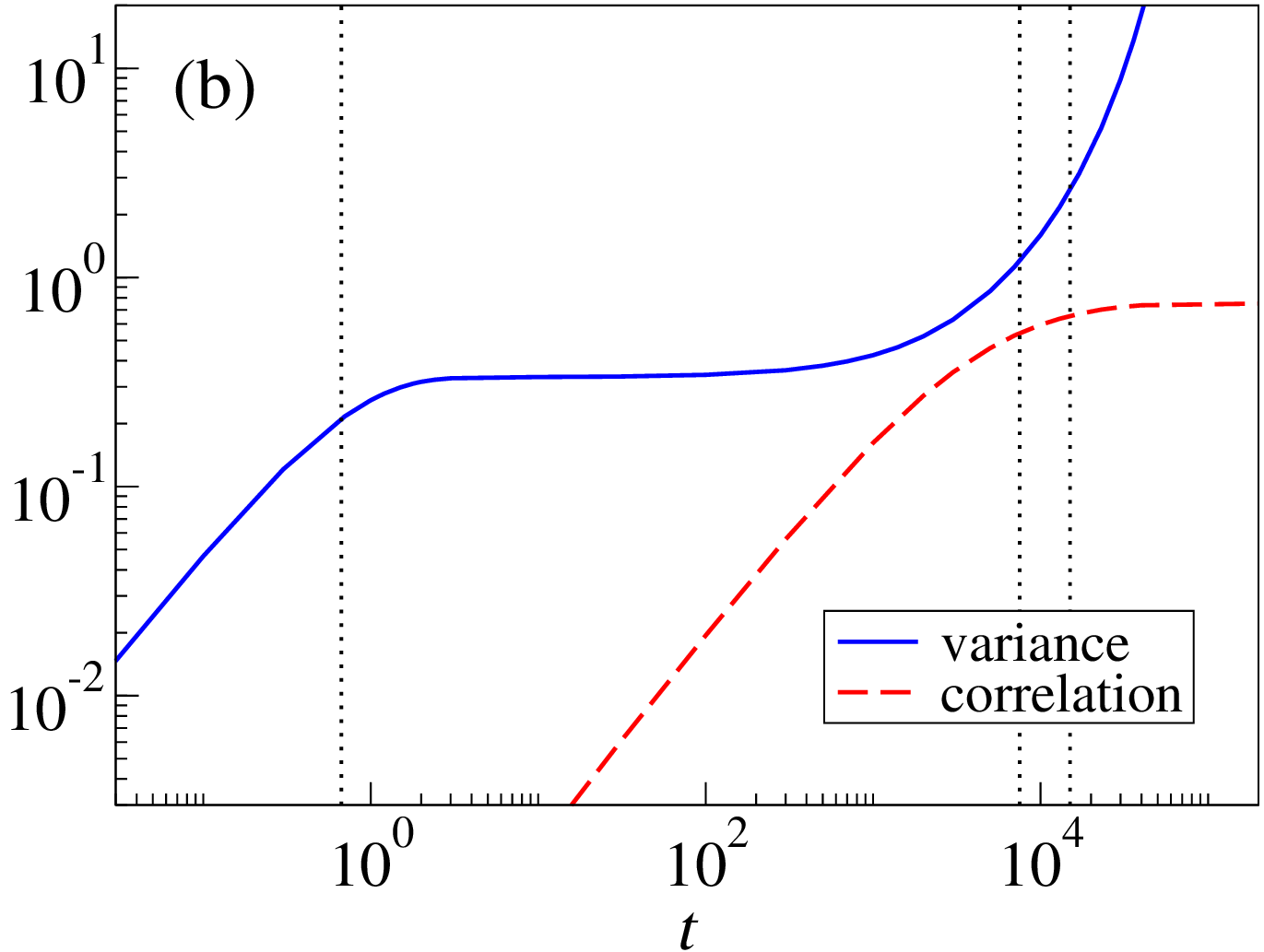}
\caption{Time evolution of $\var[v_i](t)$ and $C_{ij}(t)$ for
$10$ agents (a) and for $10^4$ agents (b). Analytical results
following from \eref{evolution-CN} are shown as lines, numerical
results obtained by averaging over $10^5$ realisations are shown
as symbols, $\sigma^2=0.5$. Vertical dotted lines indicate the
transition times $t_1$, $t_2$, and $t_3$, left to right,
respectively.}
\label{fig:time_evol}
\end{figure}

To verify the presented analytical results we investigated the
model numerically. For numerical solutions of stochastic
differential equations we used Milstein's
method~\cite{Gard04,Hig01}; random numbers were generated using
the standard GSL library and the Mersenne Twister
generator~\cite{GSL}, and the time increment was $10^{-4}$ in
all simulations. In the used discretisation scheme, there is
a~non-zero probability that the wealth $v_i(t)$ becomes
negative~\cite{Moro04}. However, thanks to the typical value of
$v_i(t)$ and the small time step, in the presented numerical
simulations this was not an issue. As can be seen
in~\Fref{fig:time_evol}a, our analytical results agree with
numerical simulations of the system. Due to the small number of
agents, transition times $t_{2,3}$ are small and the system goes
directly from the free regime to the synchronized regime.
In~\Fref{fig:time_evol}b the number of agents is large and the
system behaviour is more complex. In the initial period the
variance is small and correlations are negligible, while in the
period $t\in[2;300]$ the variance is almost constant and
correlations are still small---the system is in the power-law
regime (due to large computational complexity, no numerical
results are shown here). Eventually, for $t\gtrsim10^4$, the
synchronized regime is established. The transition times given
by~\eref{t1}, \eref{t2}, and \eref{t3} are shown as vertical
dotted lines and agree well with the described changes of the
system behaviour.

We should sound here a note of caution about the interpretation
of the averages $\avg{v_i^2(t)}$ and $\avg{v_i(t)v_j(t)}$ and the
wealth distribution $f(v_i,t)$. All these quantities are
ensemble-based: if many copies of the system evolve
independently for time $t$, by examining the final wealths of
agent $i$ one can estimate both the distribution $f(v_i,t)$ and
the averages. By contrast, when one speaks about an empirical
wealth distribution, that is based on the wealth of all agents
in one realisation only, it is population-based. However, when
the number of realisations and the number of agents are large
and the wealth correlations are small, ensemble- and
population-based quantities are alike. Such behaviour was
observable also in the numerical simulations presented above. In
the free and power-law regimes, the variance of wealth in each
realisation was similar to $\var[v_i(t)]$ (at various times,
differences were less than $20\%$ for $N=10$ and less than $1\%$
for $N=10\,000$) and its relative fluctuations between
realisations were approximately $50\%$ for $N=10$ and $2\%$ for
$N=10\,000$. As time goes on, fluctuations of the
population-based variance grow and so does the difference
between the ensemble-based and population-based variance of
wealth. In the synchronized regime, the equivalence of the two
quantities breaks entirely.

\begin{figure}
\centering
\includegraphics[scale=0.28]{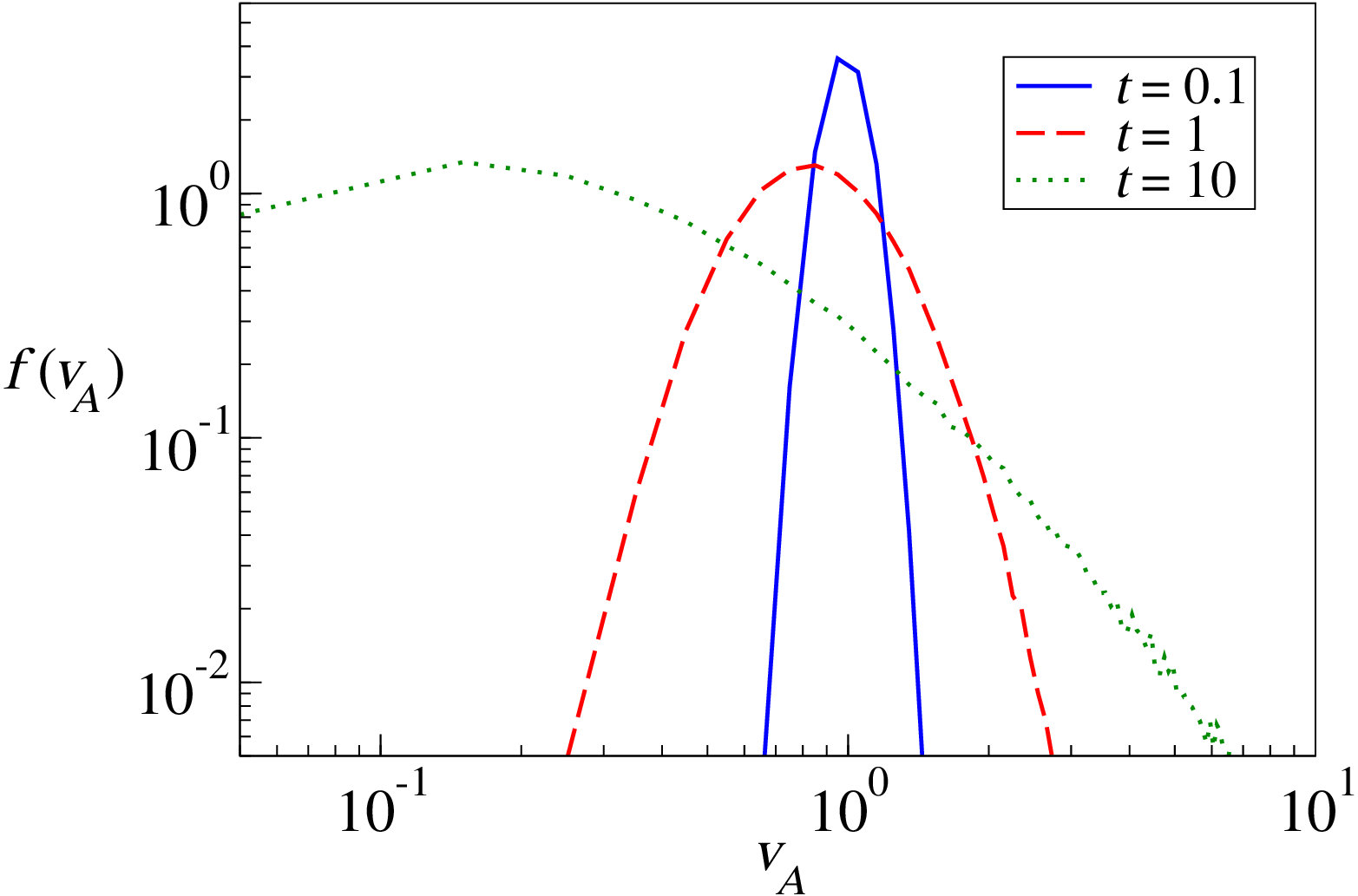}
\caption{Probability density of the average wealth $v_A$ at
various times. Parameter values are $N=10$ and $\sigma^2=0.5$,
probabilities were obtained from $10^5$ independent
realisations of the model.}
\label{fig:v_A}
\end{figure}

The nature of the synchronized regime can be better understood
by recalling the average wealth $v_A(t)$ again. As explained
above, its evolution is given by a sum of multiplicative
processes, $\dd v_A(t)=\case{\sqrt2\sigma}N
\sum_{i=1}^N v_i(t)\,\dd W_i(t)$. Despite this summation of
contributions and their variable strengths ($\avg{v_i^2(t)}$
increases with time), according to \Fref{fig:v_A} the
distribution of $v_A$ is approximately lognormal and, in
agreement with our expectations, the variance $\var[v_A]$ is
increasing. In the initial regime, this increase is due to
growing variances of all agents' wealth. In the power-law
regime, variances of wealth levels are approximately constant
but their growing correlations lead to increasing $\var[v_A]$.
In the synchronized regime, wealth correlations are already
saturated and the growth of $\var[v_A]$ is caused by
exponentially growing variances of wealths. Since correlations
are large, ensemble- and population-based quantities are no
longer equivalent. Finally we remark that since $v_A>0$,
$\avg{v_A(t)}=1$ is fixed, and $\var[v_A(t)]$ grows without
bounds, in the course of time it is increasingly probable that
$v_A(t)$ is much smaller than its expected value
$\avg{v_A(t)}=1$; this can be interpreted as a high occurrence
of temporal depressions of the economy.

When $\sigma\geq1$, both $\var[v_i](t)$ and $C_{ij}(t)$ diverge
and must be replaced by different quantities. Instead of the
variance, one can use the mean absolute deviation
$\langle\vert v_i(t)-1\vert\rangle^2$ which avoids second
moments of the wealth distribution and hence can be used for any
$\sigma$. The Pearson's correlation coefficient can be replaced
by a rank correlation coefficient (Kendall's $\tau$ or
Spearman's $\rho$). All three proposed quantities are hard to
handle in analytical calculations and with strong noise,
numerical simulations of the system are extremely
time-demanding. While we have obtained no definite results yet,
preliminary outcomes suggest that in this case too the
transition from the power-law regime occurs at a time
proportional to the number of agents $N$.

\section{General exchange network}
\begin{figure}
\centering
\includegraphics[scale=0.28]{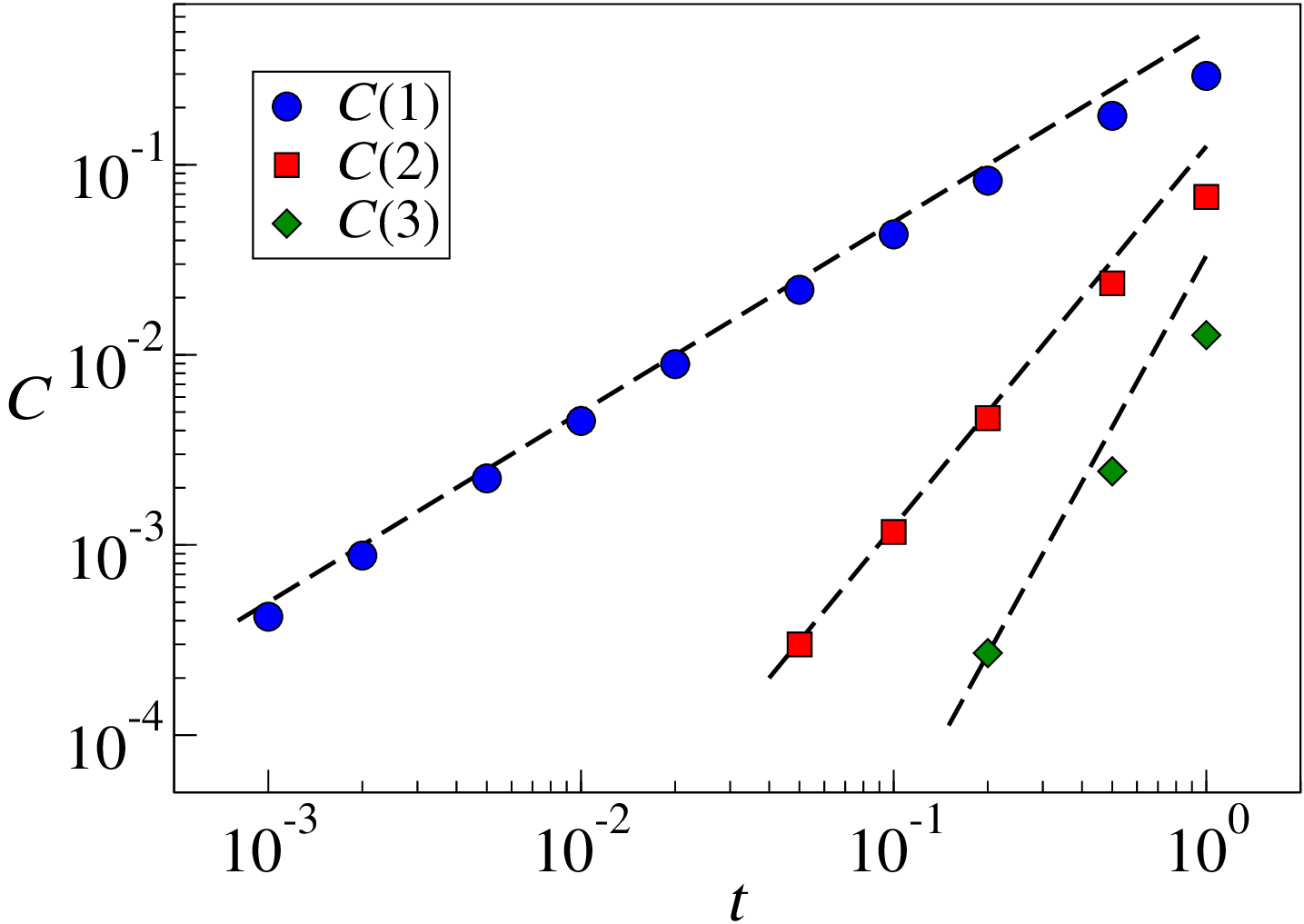}
\caption{Time evolution of correlations for the ring network of
10 agents, $\sigma^2=0.25$. Symbols show numerical results for
neighbouring agents (circles), agents with the distance 2
(squares), and agents with the distance 3 (diamonds), averaged
over $10^7$ realisations. Dashed lines have slopes 1, 2, and 3,
respectively.}
\label{fig:ring}
\end{figure}

Now we generalize the exchange network to an arbitrary graph:
denoting the set of neighbours of agent $i$ by $\N_i$, the
number of neighbours by $k_i$, the average number of neighbours
by $z$. We assume that each agent interacts equally with all
neighbours and per unit time exchanges the whole wealth, hence
\begin{equation}
\label{J-general}
J_{ij}=1/k_j\quad\mbox{for $i\in\N_j$},\quad
J_{ij}=0\quad\mbox{for $i\not\in\N_j$};
\end{equation}
notice that the matrix of exchanges $\mathsf{J}$ is asymmetric.
Now, \eref{dv-general} generalizes to
\begin{equation}
\label{dv-network}
\dd v_i=\big(\hat v_i-v_i\big)\,\dd t+\sqrt2\sigma v_i\,\dd W_i
\end{equation}
where $\hat v_i:=\sum_{j\in\N_i}v_j/k_j$. By averaging over
realisations we obtain the set of equations for the stationary
values of the average wealths
\begin{equation}
\label{v-stationary}
\avg{v_i}=\sum_{m\in\N_i}\frac{\avg{v_m}}{k_m}
\end{equation}
which is solved by $\avg{v_i}\sim k_i$. Assuming average wealth
equal to $1$, \eref{v-stationary} has the unique solution
$\avg{v_i}=k_i/z$. This means that the topology of the exchange
network is crucial for the distribution of wealth among the
agents. Consequently, when $\sigma$ is small and hence wealth
fluctuations are negligible, a power-law distribution of wealth
can be purely a topological effect of a~scale-free degree
distribution in the network of agent exchanges. To proceed,
$\avg{v_i^2(t)}$ and $\avg{v_i(t)v_j(t)}$ are again the key
quantities. They fulfill the equations
\begin{equation}
\label{genN-evolution}
\eqalign{
\frac{\dd\avg{v_i^2}}{\dd t}=&
2\sum_{m\in\N_i}\frac{\avg{v_iv_m}}{k_m}-
2(1-\sigma^2)\avg{v_i^2},\\
\frac{\dd\avg{v_iv_j}}{\dd t}=&
\sum_{m\in\N_i}\frac{\avg{v_jv_m}}{k_m}+
\sum_{n\in\N_j}\frac{\avg{v_iv_n}}{k_n}-2\avg{v_iv_j}}
\end{equation}
which can be derived similarly to \eref{evolution-CN}. We set
the initial conditions according to the stationary wealths as
$v_i(0)=k_i/z$ and thus $\avg{v_i^2(0)}=k_i^2/z^2$ and
$\avg{v_i(0)v_j(0)}=k_ik_j/z^2$ (the general case is studied in
\Sref{sec:initial}). From \eref{genN-evolution} follows
\begin{equation}
\label{genN-initial}
\left.\frac{\dd\avg{v_i^2}}{\dd t}\right\vert_{t=0}>0,\quad
\left.\frac{\dd\avg{v_iv_j}}{\dd t}\right\vert_{t=0}=0,
\end{equation}
which means that the growth of $\var[v_i](t)$ precedes the
growth of $C_{ij}(t)$. This gives us a way to investigate the
small time behaviour of \eref{genN-evolution}: assuming
$\avg{v_i(t)v_j(t)}$ constant, we obtain $\avg{v_i^2(t)}$ which
in turn leads to an enhanced estimate of $\avg{v_i(t)v_j(t)}$.
For neighbouring agents $i$ and $j$, the results are
\begin{eqnarray}
\label{genN-var-smallt}
\var[v_i(t)]&=\frac{2\sigma^2k_i^2}{z^2}\,t+O(t^2),\\
\label{genN-C-smallt}
C_{ij}(t)&=\frac{k_i+k_j}{2k_ik_j}\,t+O(t^2).
\end{eqnarray}
Moreover, it can be shown that when the shortest path between
agents $i$ and $j$ has the length $L$, the leading term of
$C_{ij}(t)$ is proportional to $t^L$. These results are
confirmed by \Fref{fig:ring} where we investigate a system of
ten agents who are placed on a ring (\emph{i.e.}, $k_i=2$,
$i=1,\dots,10$). As can be seen, numerical results agree well
with $C_{ij}(t)$ proportional to $t^L$. The system produces
a~``cascade'' of correlations: first only neighbouring agents
are considerably correlated, then also agents with the distance
two, distance three, and so forth.

For \eref{dv-network}, the mean-field approximation yields the
stationary distribution
\begin{equation}
\label{genN-mean-field}
f_i(v_i)=K_i\exp[-(\lambda_i'-1)/v_i]v_i^{-2-\lambda_i'}
\end{equation}
where $\lambda_i'=1+k_i/(z\sigma^2)$ and the corresponding
variance is $\var[v_i]=k_i^2\sigma^2/(z^2(1-\sigma^2))$. By
comparing this stationary variance with \eref{genN-var-smallt},
we obtain the transition time from the free regime to the
power-law regime as
\begin{equation}
\label{genN-t1}
t_1'=\frac1{2(1-\sigma^2)}
\end{equation}
which is identical to \eref{t1}. Further, from
\eref{genN-C-smallt} we see that the transition time from the
power-law regime to the synchronized regime is proportional to
$k_ik_j/(k_i+k_j)$ and thus for the whole network it can be
estimated as
\begin{equation}
\label{genN-t2}
t_2'=O(z)
\end{equation}
which is a generalization of \eref{t2}. We see that for networks
with a relatively small average degree, the power-law regime
appears only for a limited time or not at all.

We were unable to obtain an equivalent of the transition time
$t_3$ for a~general network. Considering, for example, a simple
star-like structure with one agent in the center and the
remaining $N-1$ agents connected only to him, one can see that
the transition time $t_3'$ is small and does not scale with $N$.
This suggests that similarly to $t_2'$, $t_3'$ is also of the
order $O(z)$. This contradicts the findings presented
in~\cite{BM00} (page~541) where they report stationary power-law
tails for $z=4$; it is possible that their numerical results are
influenced by finite-time and finite-size effects.

\subsection{Influence of the initial conditions}
\label{sec:initial}
There is still one more transition time to investigate. When the
initial conditions $v_i(0)$ are not set in line with the
stationary wealths given by Eq.~\eref{v-stationary}, a certain
time is needed to redistribute the excessive wealth levels over
the network; we say that the system is in the
\emph{equilibration regime}. Since $\avg{\dd W_i}=0$, noise
terms do not contribute to the redistribution. Thus,
\eref{dv-network} effectively simplifies to
$\dd v_i=(\hat v_i-v_i)\,\dd t$ which leads to the exponential
convergence of $v_i$ to the stationary value $k_i/z$. By the
substitution $u_i:=v_i-k_i/z$ we obtain
\begin{equation}
\label{equilibration}
\dot u_i+u_i-\sum_{j\in\N_i}\frac{u_j}{k_j}=0
\end{equation}
whose time scale is given by the initial terms $\dot u_i+u_i$ as
$O(1)$. Thus, the initial wealth distribution equilibrates in
time $O(1)$. Since the transition from the free regime occurs
roughly at the same time, the system passes from the
equilibration regime directly to the power-law regime.

\section{Conclusion}
We have shown that in the investigated model, agent wealths have
no stationary distribution and the power-law tailed distribution
reported in previous works is only transient. In addition, for
any finite number of agents, their average wealth $v_A$ follows
a multiplicative process with a fixed expected value $\avg{v_A}$
and an increasing variance $\var[v_A]$. Hence, as illustrated in
\Fref{fig:v_A}, the probability $P(v_A<x)$ approaches $1$ for
any $x>0$. We can conclude that the simple economy produced by
the model is an uneasy one: the longer it evolves, the higher
the probability that a given agent $i$ has wealth much smaller
than any positive fraction of the expected wealth $\avg{v_i}$.

There is also a more general lesson to be learned. In essence,
the mean-field approximation here anchors the agent wealths to
their expected values and thus weakens the diffusive nature of
the studied stochastic system. Mathematically speaking, the
system behaviour depends on the order of limits $N\to\infty$ and
$t\to\infty$: in the former case there is a stationary wealth
distribution, in the latter case there is none. This is an
undesired consequence of the mean-field approximation which,
as with other stochastic models, should be used with great
caution. In particular, when using it, one should check if the
nature of the studied system is not changed. To achieve this, in
this paper we have used an aggregate quantity (the average
wealth) and a quantity obtained using the mean-field
approximation (the wealth variance).

On the other hand, in some cases an anchoring term may be
appropriate. For example, a~simple taxation of wealth can be
achieved by introducing the term $r(1-v_i)\,\dd t$ to
\eref{dv-homo}, where $r>0$ represents the tax rate. Then the
set of equations for $\avg{v_i^2(t)}$ and $\avg{v_i(t)v_j(t)}$
has a nontrivial stationary solution for $\sigma<1$; one can say
that the proposed taxation stabilizes the system. Notably,
systems of coupled stochastic equations with
multiplicative noise and negative feedback are common in the
study of nonequilibrium phase transitions in magnetic
systems~\cite{Birner02}. Our work shows that this negative is
crucial for mean-field studies of such systems~\cite{Munoz05}.

In addition to the presented results, several questions remain
open. First, for large time $t$, the analytical form of the
wealth distribution $f(v_i,t)$ is unknown. Second, for an
arbitrary network of exchanges, the limiting value of the
correlation $C_{ij}(t)$ and also the transition time $t_3'$ are
of interest. Third, the strong noise case deserves more
attention and perhaps an attempt for approximate analytical
results. Finally, the studied model is simplistic, since it
combines two ingredients of economy---trade and speculation---in
a~very unrealistic way. Devising a more adequate model remains
a~future challenge.

\ack
We acknowledge the hospitality of the Comenius University
(Bratislava, Slovakia) and the Fribourg University (Fribourg,
Switzerland). We thank Franti\v sek Slanina for early
discussions and Zolt\'an Kuscsik and Joseph Wakeling for helpful
comments.


\section*{References}

\begin{thebibliography}{99}
\bibitem{Pareto} Pareto V 1897
\emph{Cours d'economie politique} (Lausanne: Rouge)

\bibitem{Pig84} Piggott J 1984
\emph{Economic Record} \textbf{60} 252--265

\bibitem{Ao00} Aoyama H et al 2000
\emph{Fractals} \textbf{8} 293--300

\bibitem{DrYak01} Dr\u agulescu A and Yakovenko V M 2001
\emph{Physica A} \textbf{299} 213--221

\bibitem{Sinha06} Sinha S 2006
\emph{Physica A} \textbf{359} 555--562

\bibitem{Champ53} Champernowne D G 1953
\emph{The Economic Journal} \textbf{63} 318--351

\bibitem{WW57} Wold H O A and Whittle P 1957
\emph{Econometrica} \textbf{25} 591--595

\bibitem{Stig69} Stiglitz J E 1969
\emph{Econometrica} \textbf{37} 382--397

\bibitem{Isp98} Ispolatov S et al 1998
\emph{Eur. Phys. J. B} \textbf{2} 267--276

\bibitem{DrYak00} Dr\u agulescu A and Yakovenko V M 2000
\emph{Eur. Phys. J. B} \textbf{17} 723--729

\bibitem{Sla04} Slanina F 2004
\emph{Phys. Rev. E} \textbf{69} 046102

\bibitem{Pat06} Patriarca M et al 2006
The ABCD's of statistical many-agent economy models
\emph{Preprint} arXiv:physics/0611245

\bibitem{ChCh07} Chatterjee A and Chakrabarti B K 2007
\emph{Eur. Phys. J. B} \textbf{60} 135--149

\bibitem{BM00} Bouchaud J-P and M\'ezard M 2000
\emph{Physica A} \textbf{282} 536--545

\bibitem{Sorin01} Solomon S and Richmond P 2001
\emph{Physica A} \textbf{299} 188--197

\bibitem{MAH04} Di Matteo T et al 2004
in \emph{The Physics of Complex Systems (New Advances and
Perspectives)} Eds. Mallamace F and Stanley H E
(Amsterdam: IOS Press)

\bibitem{Sorn98} Sornette D 1998
\emph{Phys. Rev. E} \textbf{57} 4811--4813

\bibitem{HuSo01} Huang Z-F and Solomon S 2001
\emph{Physica A} \textbf{294} 503--513

\bibitem{Reed01} Reed W J 2001
\emph{Economics Letters} \textbf{74} 15--19

\bibitem{QRR97} Quadrini V and R\'ios-Rull J-V 1997
Models of the distribution of wealth
Research Department, Federal Reserve Bank of Minneapolis

\bibitem{DS00} Davies J B and Shorrocks A F 2000
The Distribution of Wealth
\emph{Handbook of Income Distribution (Handbooks in Economics)}
Eds. Atkinson A B and Bourguignon F (Amsterdam: North Holland)

\bibitem{Yak07} Yakovenko V M 2007
Statistical Mechanics Approach to Econophysics
\emph{Springer Encyclopedia of Complexity and System Science}
(Springer Verlag)

\bibitem{New05} Newman M E J 2005
\emph{Contemporary Physics} \textbf{46} 323--351

\bibitem{Farm06} Farmer J D and Geanakoplos J 2006
Power laws in economics and elsewhere
\emph{SFI Technical report}

\bibitem{Sou01} Souma W et al 2001
Small-World Effects in Wealth Distribution
\emph{Preprint} cond-mat/0108482

\bibitem{GaLo04} Garlaschelli D and Loffredo M I 2004
\emph{Physica A} \textbf{338} 113--118

\bibitem{GaLo08} Garlaschelli D and Loffredo M I 2008
\emph{J. Phys. A: Math. Theor.} \textbf{41} 224018

\bibitem{Gup08} Gupta A K 2008
\emph{Physica A} \textbf{387} 6819--6824

\bibitem{Gard04} Gardiner C W 2004
\emph{Handbook of Stochastic Methods, 3rd Edition}
(Berlin: Springer)

\bibitem{VanK07} Van Kampen N G 2007
\emph{Stochastic Processes in Physics and Chemistry, 3rd Edition}
(Amsterdam: North Holland)

\bibitem{Hig01} Higham D J 2001
\emph{SIAM REVIEW} \textbf{43} 525--546

\bibitem{GSL} Galasi M et al 2002
\emph{GNU Scientific Library (Reference Manual)}
(Bristol: Network Theory Ltd)

\bibitem{Moro04} Moro E 2004
\emph{Phys. Rev. E} \textbf{70} 045102(R)

\bibitem{Birner02} Birner T et al 2002
\emph{Phys. Rev. E} \textbf{65} 046110

\bibitem{Munoz05} Mu\~noz MA et al 2005
\emph{Phys. Rev. E} \textbf{72} 056102
\end{thebibliography}
\end{document}